\documentclass[conference]{IEEEtran}
\IEEEoverridecommandlockouts

\usepackage[utf8]{inputenc}
\usepackage{silence}
\usepackage{longtable}
\usepackage{colortbl}
\usepackage{cite}
\usepackage{amsmath,amssymb,amsfonts}
\usepackage{algorithmic}
\usepackage{graphicx}
\usepackage{textcomp}
\usepackage{amsthm}
\usepackage{amsmath}
\usepackage[left=1.2cm, right=1cm, top=1.6cm, bottom=1.2cm]{geometry}

\usepackage{amssymb}
\usepackage{bm}
\usepackage{bbm}
\usepackage{subfigure}
\usepackage{setspace}
\usepackage{graphicx}
\usepackage{calc}
\usepackage{epstopdf}

\usepackage{verbatim}
\usepackage{color}
\usepackage[dvipsnames]{xcolor}
\usepackage{array}
\usepackage{cite}
\usepackage{xspace}
\definecolor{ForestGreen}{RGB}{34,139,34}
\usepackage{algorithm}
\usepackage{multicol}

\usepackage{caption}
\def\BibTeX{{\rm B\kern-.05em{\sc i\kern-.025em b}\kern-.08em
            T\kern-.1667em\lower.7ex\hbox{E}\kern-.125emX}}

\usepackage{newunicodechar}
\DeclareUnicodeCharacter{FB01}{fi} 
\DeclareUnicodeCharacter{FB02}{fl} 
\DeclareUnicodeCharacter{2061}{}

\begin{document}
	
\title{A Robust Clustering Scheme for Vehicular Communication Networks}

\author{\IEEEauthorblockN{ Maryam Hosseini}
		\IEEEauthorblockA{\textit{Department of Electrical Engineering} \\
		\textit{Polytechnique Montréal}\\
		Montréal, Canada \\
		maryam.hosseini@polymtl.ca}
	\and
	\IEEEauthorblockN{ Gunes Karabulut Kurt}
	\IEEEauthorblockA{\textit{Department of Electrical Engineering} \\
		\textit{Polytechnique Montréal}\\
		Montréal, Canada \\
		gunes.kurt@polymtl.ca}
}

\maketitle
	
	\begin{abstract}
		
		Clustering, as a technique for grouping nodes in geographical proximity together, in vehicular communication networks, is a key technique to enhance network robustness and scalability despite challenges such as mobility and routing. This paper presents a robust clustering scheme based on cluster head backup list algorithm for unmanned aerial vehicles (UAVs)-assisted vehicular communication network, where multiple UAVs act as communication base stations for a vehicular network. To tackle the high mobility issues in vehicular communications, instead of allowing direct communication between all vehicles to the UAV, clustering methods will potentially be efficient in overcoming delay limitations, excessive power consumption and resource issues. Using the clustering technique, neighboring vehicles are grouped into clusters with a specific vehicle selected as the cluster head (CH) in each cluster. The selected CH connects directly to the UAV through an infrastructure-to-vehicle (I2V) link, subsequently establishing vehicle-to-vehicle (V2V) communications with vehicles in the same cluster. To increase cluster connectivity period, the proposed clustering scheme is developed based on considering the vehicle behavior for efficient selection of CHs and providing a CH backup list to maintain the stability of the cluster structure. Numerical evaluations show that the proposed system outperforms benchmark schemes in terms of clustering stability and reliability. It is also shown that the performance of the proposed scheme is not much affected by the increase in the number of vehicles. This indicates that the proposed scheme can be efficient in dense vehicular networks where resource constraints pose significant challenges.
	\end{abstract}

	\IEEEpeerreviewmaketitle
	
	\begin{figure*}[t!]
		\begin{center}
			\includegraphics[width=200 mm, height=75 mm, keepaspectratio]{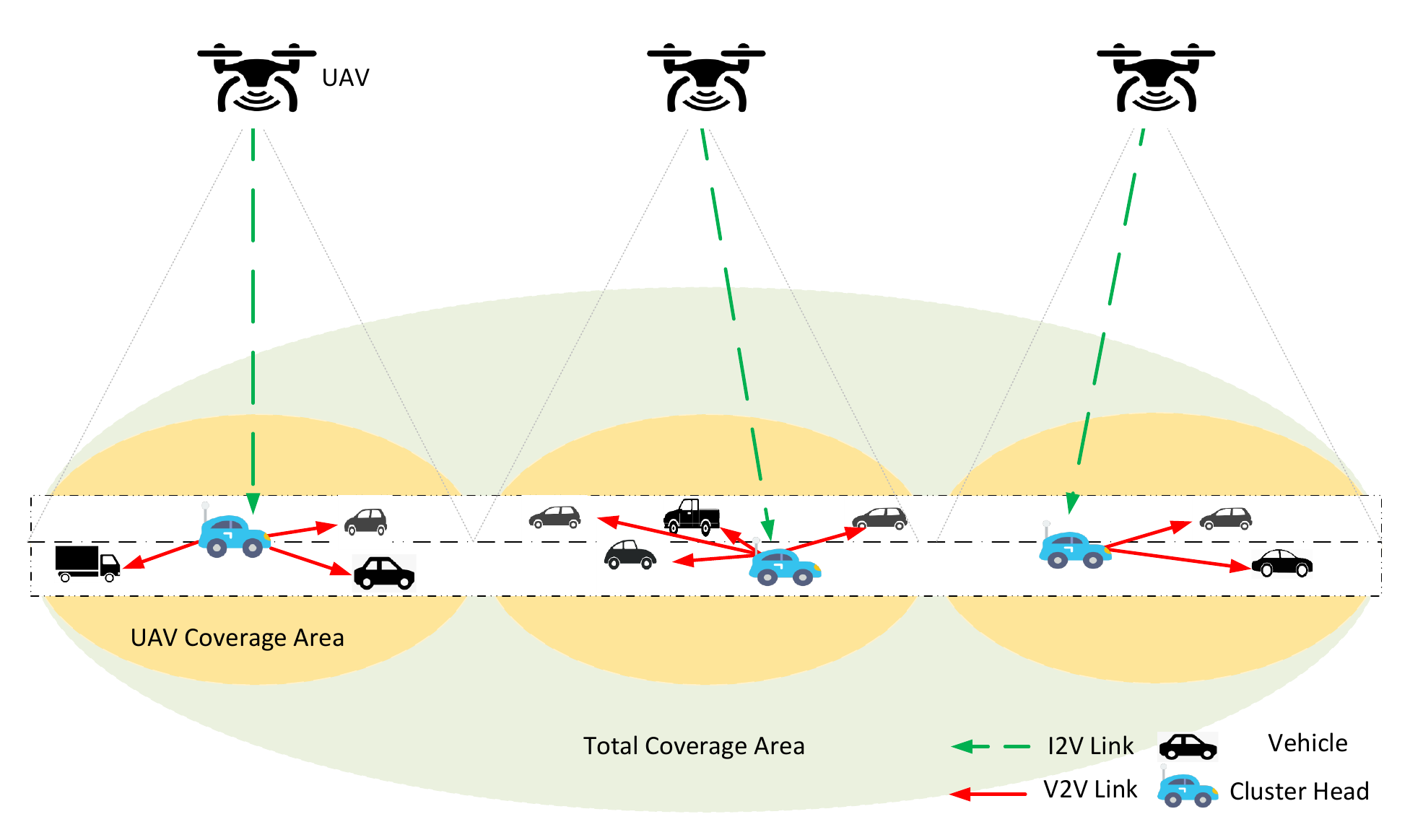} 
   \vspace{-2mm}
			\caption{System Model.}
			\label{System-model-1}
		\end{center}
	\end{figure*}
	
	\section{INTRODUCTION}
	The increasing demand for mobile data traffic on the road has spurred significant interest in vehicular communication networks, deploying infrastructure-to-vehicle (I2V) and vehicle-to-vehicle (V2V) communication links \cite{1,mukhtaruzzaman2020clustering,hosseini2022stackelberg,47,R403}. These networks offer key benefits such as increased safety, connectivity, and the provision of in-flight entertainment for passengers, including online gaming and video streaming. Achieving these goals requires reliable communication between vehicles and cellular infrastructure. One proposed approach to enhance connectivity in vehicular networks is for selected vehicles to act as mobile base stations (BSs), with trucks and buses as mobile femtocell access points (FAPs) to support vehicles in their coverage areas, as suggested by \cite{11}. However, while this approach shows efficiency in scenarios with mobile FAPs, in many cases due to the high mobility of vehicular networks, the availability of routes remains uncertain and is a challenge for cellular infrastructure \cite{R402,ozcan2021reconfigurable}. A potential solution is to use unmanned aerial vehicles (UAVs) or drones as aerial BSs or relays \cite{R3}.
	
	Due to their mobility, height adjustment, on-demand deployment, and cost-effectiveness, UAVs have attracted attention and offer a promising technology for improving communications in terrestrial wireless networks \cite{al2017modeling,arafat2018survey,hosseini2022game}. In \cite{36}, researchers investigated a UAV-enabled multicast channel where a UAV acts as an aerial transmitter for a group of ground users and analyzed the channel capacity under power and speed constraints. Using UAV mobility and flexible deployment, it can provide robust I2V links, improve channel quality, and line-of-sight (LoS) links to support vehicle communications, especially in scenarios with hazardous environments or temporary traffic disruptions where ground infrastructure is unavailable \cite{R1,20,21,49,R2}. In addition, UAV deployment reduces reliance on vehicle congestion and inter-vehicle cooperation, increases link path availability, and reduces packet delivery delays \cite{16}. Despite the advantages of using a UAV as an aerial base station, there are still challenges, mainly related to energy and limited coverage range. To deal with this problem, deploying multiple UAVs dynamically as a network increases capabilities compared to a single UAV, thereby improving system performance \cite{41,56, li2021energy}. However, optimization of trajectory design, coverage, and resource allocation for multi-UAV networks are still fundamental challenges.\\
	\indent	Considering the specific movement patterns of the vehicle, clustering is considered one of the most important techniques used to increase the stability, reliability, and scalability of vehicular networks. In this context, proposing a robust clustering scheme, where a cluster head (CH) is effectively selected, can increase the scalability and reliability of multi-UAV-assisted vehicular networks \cite{zhang2018new}. Thereby, instead of direct communications between UAVs and all connected vehicles, which requires significant energy and time resources, the selected CH can effectively manage intra-cluster communication. This approach leads to reduced energy consumption and minimized latency and increases overall network efficiency. For instance, \cite{inam2019novel} proposed the use of K-means and Floyd-Warshall algorithms to form vehicle clusters and then select the vehicle head as CH based on the optimal route to the closest vehicle and showed that the proposed scheme increases the stability of the cluster.\\
		\indent	Please note that due to the recent developments in vehicular communication, the European Telecommunications Standards Institute (ETSI) has assigned a specific form of broadcast communication referred to as Cooperative Awareness Messages (CAMs) within the realm of vehicular communications. These messages contain information crucial to safety and stability-related applications, encompassing details such as vehicle speed, position, and direction \cite{etsi2014302}. 
	
	\subsection{Motivation and Contributions}
 \vspace{-2mm}
	\indent This paper presents a framework for employing a reliable clustering scheme with a novel CH selection algorithm in a UAV-assisted vehicular communication network. The scenario involves multiple UAVs acting as communication base stations to serve a vehicular network. With an efficient CH selection scheme, V2V links between the selected CHs and vehicles can be established instead of allowing direct communication between all vehicles to UAVs (I2V links), to overcome energy limitations and tackle the high mobility issues in vehicular communications. 
	
	\indent In this regard, initially, a clustering scheme is executed, in which every vehicle on the road is assigned to a specific UAV. Following that, the UAV collects all CAMs from the assigned vehicles to select the most reliable vehicle as a CH. Then, the UAV directly communicates the CH, and subsequently, the CH bears the responsibility for the communications with other vehicles within the cluster. To increase the cluster connectivity time, a CH backup list-providing algorithm is proposed to maintain the stability of cluster structures. The main contributions of this paper can be outlined as follows:
	\begin{itemize}
		\item Employing UAVs can overcome the lack of ground base stations by establishing LoS links.
		\item Instead of allowing direct communications between each vehicle and UAVs which leads to the communication and scheduling overhead at UAVs, an efficient clustering is proposed in which neighboring vehicles are grouped and are connected to a particular CH instead of the UAV. 
		\item To more efficient communications within clusters, a creative CH selection based on the knowledge of vehicle behavior such as its average speed, position, direction, and the number of neighbors is proposed. Considering these parameters can increase stability and connectivity and reduce additional costs in the network.
		\item By providing a CH backup list, the fast selection of an alternate CH by the UAV is enabled, effectively mitigating the challenges caused by the dynamic nature of vehicular communication and promoting cluster connectivity.
	\end{itemize}
	
	\subsection{Paper Organization}
	The subsequent sections of this paper are structured as follows: In Section II, the system model of a UAV-assisted vehicular communication network is presented. Section III introduces a novel clustering scheme including user assignment, CH selection, and providing a CH backup list. Finally, Section V is dedicated to addressing open issues and conclusions.

	\section{System Model}
	\subsection{Spatial Model}
	According to Fig. 1, the considered vehicular communication network consists of $ I $ vehicles equipped with
	omnidirectional antennas, traveling on a straight two-line road, and $ J $ UAVs equipped
	with directional antenna. All UAVs fly at a maximum speed $v_{max}$ at a fixed altitude $h_U$ to support the vehicular communication networks on the road via air-to-ground (A2G) channels. A minimum distance between the UAVs, denoted by $ d_{\text{min}} $, is considered to both avoid collisions and ensure coverage of the entire area. The vehicles and UAVs sets are denoted as $ \mathcal{I}=\{1,2,...,I\} $ and $ \mathcal{J}=\{1,2,...,J\} $, respectively, where $ |\mathcal{I}| = I $ and $ |\mathcal{J}| = J $.\\
	\indent Let us consider that the total process time, $ T $, is split into $ N_t $ equal time slots with the duration of $ \Delta t = \frac{T}{N_t} $. We assume that the two-dimensional (2D) Cartesian coordinate of each vehicle $ V_i $ is represented as $ {\bf{q}}_{V_i}(t) = [x_i(t),y_i(t)]^T \in R^{2\times 1}, i \in I $. Moreover, considering all UAVs are flying at a ﬁxed altitude from the road surface, the time-varying 3D Cartesian coordinate of the $ j $th UAV at time instant $ t $, $ 0 \leq t \leq \Delta t $, can be expressed as ${\bf{q}}_{U_j}(t)= [x_j(t),y_j(t),h_U]^T \in R^{3\times 1}, j \in J $. Therefore, the Euclidean distance from the $ j $th UAV to the $ i $th vehicle at a time instant $ t $, $ 0 \leq t \leq \Delta t $, can be expressed as:
	\begin{equation}
		\begin{aligned}
			d_{i,j}(t)&=\lvert\lvert{{\bf{q}}_{U_j}(t)-{\bf{q}}_{V_i}(t)}\rvert\rvert\\&=\big((x_{V_i}(t)-x_{j}(t))^2+(y_{V_i}(t)-y_{j}(t))^2+h_U^2\big)^{\frac{1}{2}}.
		\end{aligned}
	\end{equation}
	
	We adopt the Orthogonal Frequency Division Multiple Access (OFDMA) technique, ensuring that all links within the network have orthogonal access to the bandwidth. This setup guarantees the absence of intra-network interference, thus promoting efficient and interference-free communication among network nodes.
	
	\subsection{Communication Model}
	\indent The UAVs with the capability of flexible altitude adjustment can offer a perfect LoS link for ground communication networks. For simplicity in channel modeling, according to \cite{31}, we assume that there is no small-scale fading for I2V links. It is also assumed that all I2V links are LoS dominated and multipath fading can be ignored \cite{49}. Hence, the quality of the I2V link mainly depends on the Euclidean distances from UAVs to vehicles. As a result, the channel gain from the $ i $-th UAV to the $ j $-th vehicle at the time instant $ t $, corresponding to the free space path loss model, can be expressed as $g_{i,j}(t)=g_0d_{i,j}^{-2}(t)$, where $ g_0 $ expresses the channel gain at the reference distance $ d_0 = 1 $ $m $, defined by the International Telecommunication Union (ITU) \cite{31, 49}.
	
	Let us assume that the binary parameter $ \delta_{i,j} $ represents the UAV-vehicle assignment. In this regard, $ \delta_{i,j}= 1 $ indicates vehicle $ i $ is connected to UAV $ j $, and $\delta_{i,j}= 0 $ if otherwise. Subsequently, the signal-to-noise ratio (SNR) at the $i$th vehicle from the $j$th UAV, can be obtained as follows
	
	\begin{equation}
		\gamma_{{i,j}}(t)={\delta_{i,j}(t)}\frac{P_{U_j}(t) g_{i,j}(t)}{\sigma_w},
	\end{equation}
	
	\noindent where $\sigma_{w}$ represents the noise power which is considered to be constant throughout the system, and $P_{U_j}$ states the transmission power of $j$th UAV.
	
	\indent We also define $ g_{V_{i,i'} }(t)$ as the channel gain of V2V link between $i$ and $i'$th vehicles at the time $ t $. Therefore, the channel gain in a short time slot can be obtained as
	
	\begin{equation}
		g_{V_{i,i'} }(t)=f_{V_{i,i'}}\mu_{V_{i,i'}}Ld_{V_{i,i'}}^{-\eta},
	\end{equation}
	
	\noindent where $ f_{V_{i,i'}} $ is the fast fading factor with an exponential distribution with the parameter of 1, $ \mu_{V_{i,i'}} $ states the stochastic parameter of shadowing with a $ \log $ normal distribution, $ L $ refers to the path loss constant, $ d_{V_{i,i'}} $ expresses the Euclidean distance between two vehicles, and $ \eta $ states the path loss exponent factor. We define $ J_{V_{i,i'}}=\mu_{V_{i,i'}}Ld_{V_{i,i'}}^{-\eta} $ as the large scale fading effect between two vehicles.\\
	\indent Furthermore, we assume that the Cartesian coordinates of vehicles are known a priory for a central entity. All UAVs remain connected to the central entity via wireless back-haul such as millimeter-wave (mmW) and are aware of the locations of vehicles and the result of the proposed algorithms. Moreover, all vehicles are assumed to be equipped with the on-board data unit (OBU) \cite{singh2019tutorial}.
	
		\begin{figure}[t]
		\centering
		\includegraphics[width=90 mm, height=30 mm, keepaspectratio]{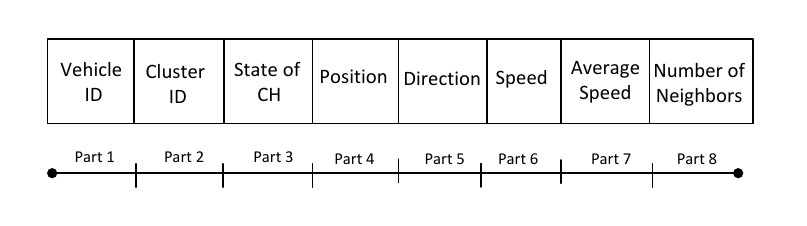} 
		\caption{CAM Structure.}
		\label{System-model-2}
	\end{figure}
	\vspace{-5mm}
	\section{Proposed Stable Clustering Scheme}
	
	The initial step of the proposed scheme involves an efficient clustering scheme, where all vehicles are clustered through a fast algorithm, followed by providing a reliable CH selection and then, an efficient scheme to provide a list of potential CH candidates aimed at bolstering the stability of the clusters.
	
	\begin{algorithm}  [t]
		\caption{ UAV-Vehicle Assignment } 
		{	Inputs: $\{d_{i,j}\}, \mathcal{I}~=~\{1,2,..,I\}, \mathcal{J}~=~\{1,2,..,J\}, \{N_j\}~=~0,   \newline \delta_{i,j}~=~0$, \indent  $\forall i,j$ ,\\
			Outputs: $\{{\bf{A}}_{j}\}=\{\delta_{i,j}\}$.
		} 
		\begin{algorithmic}[1] 
			
			\STATE {\bf{for}} $ j \in \mathcal{J} $ 
			\STATE Broadcast a pilot signal with power $P_{U_{j}}$ 
			\STATE {\bf{end} \bf{for}}
			\STATE  {\bf{for}} $ i \in \mathcal{I} $
			\STATE Compute the $ \{\gamma_{{i,j}}\} $ based on (2)
			\STATE {\bf{while}} $\mathcal{I}=\{\} $
			\STATE Find the $\arg(\max_{j} \{\gamma_{{i,j}}\})$
			\STATE Assign $ i $th vehicle to the $ j $th UAV and set $ \delta_{i,j} =1, $
			\STATE Set $ N_j=N_j+1 $
			\STATE Set $\mathcal{I}=\mathcal{I}-\{i\} $  
			\STATE {\bf{end} \bf{while}}  
			\STATE {\bf{end} \bf{for}}  
		\end{algorithmic}
	\end{algorithm}

	\subsection{Clustering Phase}
	During this phase, the vehicle-UAV association is established using a fast clustering approach outlined in Algorithm 1 \cite{hosseini2022stackelberg}. Within this algorithm, during each time period, the positions of UAVs are considered as the cluster centers. Consequently, vehicles are assigned to UAVs based on the maximum received SNR. This iterative process continues until all vehicles have been successfully assigned to UAVs. Subsequently, each UAV is assigned a cluster of vehicles requiring communications services.\\
	\indent While there are many other efficient clustering schemes such as the k-means algorithm \cite{sinaga2020unsupervised}, dealing with the mobility nature of vehicular networks is challenging in using such algorithms. Meanwhile, the proposed algorithm is not only fast, but also ensures the minimum requirement for reliable communication through an A2G link.

	\subsection{Cluster Head Selection Scheme}
	Following the execution of the clustering scheme, each vehicle within the cluster at regular intervals broadcasts a CAM in the form of Fig. 2 \cite{rossi2017stable}. This CAM is organized in 8 parts, all of which are numbered for simplicity, and are explained as follows: Parts 1 and 2 refer to the vehicle and cluster identity, $ID_i, ID_C$, respectively. Part 3 indicates the CH status, a binary parameter denoted as $\zeta_{C,i}$, where $\zeta_{C,i}=1$ if vehicle $i$ acts as a CH and $\zeta_{C,i}=0$ if otherwise. 
	Parts 4 and 5 state the 2D position and direction of the vehicle, denoted as ${\bf{q}}_{V_i}$ and $ {\boldsymbol{\theta}}_i$, respectively. Furthermore, part 6 indicates the vehicle's velocity and part 7 represents the vehicle's average speed, which can be calculated as follows
	
	\begin{equation}
		v_{avg_i}=\frac{v_i(t)+\sum_{\tau=1}^{T_w-1}v_{i}(t-\tau)}{T_w},
	\end{equation}
	
	\noindent where $v_i(t)$ denotes the velocity of the vehicle at the specific time moment $t$, and $T_w$ refers to the considered window size for computing average
	speed \cite{taha2018lightweight}. Moreover, part 8 represents the set of all neighbor
	vehicles within range of vehicle $i$ denoted by $\eta_i$. 
	
	Based on the above discussion, every UAV collects all CAMs from the connected vehicles and subsequently initiates the process of CH selection. It first computes the cluster average speed as $v_{cl_j}=\frac{\sum_{i'=1}^{N_j}v_{avg_{i'}}}{N_j}$,  where $N_j$ represents the total number of vehicles assigned to the $j$th UAV's cluster.\\
	\indent Subsequently, the UAV opts to select a vehicle to act as the cluster head. This selection considering all parameters outlined in the CAM, is mainly based on factors such as the vehicle's average speed, which should closely match the cluster's average speed, along with having a minimum number of neighbors. Additionally, the decision considers the duration a vehicle remains inside the cluster, referred to as the residual path and denoted as $\xi_i$, which can be calculated using \textit{i}th vehicle's position and direction. The proposed CH selecting approach is depicted in Algorithm 2.
	
	
	\begin{algorithm} [t]
		\caption{CH Selection Scheme} 
		{	Inputs: $\{{\bf{N}}_{j}\}, \{{\bf{q}}_{U_j}\}, \{\text{CAM}_i\}, \forall i, j $ \\
			Outputs: $\{CH_{j}\}, j\in \mathcal{J}$, where $CH_{j}$ is the selected CH for $j$th UAV cluster.
		} 
		\begin{algorithmic}[1] 
			\STATE  {\bf{for}} $ j \in \mathcal{J} $ 
			\STATE  {\bf{for}} $ i \in {\bf{N}}_j  $ 
			\STATE  Compute the cluster average speed as (4)		
			\STATE  Compute  $ v_{d_i}=\lvert\lvert v_{avg_{i}}- v_{cl_j} \lvert\lvert$ 	
			\STATE Find the $i'$th vehicle as $\arg(\min_{i} \{ v_{d_{i}}\})$		
			\STATE Set 	$\{{\bf{N}}_{j}\}-i'$	
			\STATE Compute the residual path within the cluster for $i'$ based on its position, direction and speed as $\xi_{i'}=2r_U-{v}_{avg_{i'}}\Delta t$
			
			\STATE {\bf{if}}  $\xi_{i'} \geq \epsilon_d$, and  $\eta_{i'} \geq \epsilon_n$  {\bf{then}} 
			\STATE 	Set $i'$ as the $CH_j$
			\STATE  {\bf{else}}
			\STATE Go to Step 5
			\STATE {\bf{end} \bf{if}}	
			\STATE {\bf{end} \bf{for}}
			\STATE {\bf{end} \bf{for}}
			
		\end{algorithmic}
	\end{algorithm}

	\begin{algorithm} [t]
		\caption{CH Backup List Selection Using AHP} 
		{	Inputs: List of remaining vehicles from Alg. 2 $\{\bf{\bar{N}}\}_j$, residual path set calculated in Alg. 2 $\{{\boldsymbol{\xi}}_i\}$, $ \{\text{CAM}_i\}$, $\{ {\bf{CB}}_j\}=\{\} $, $\{{\bf{v}}_{d_i}\}$, $ \forall i, j $ \\
			Outputs:  CH-backup list, $\{ {\bf{CB}}_j\} $, $ \forall  j $.} 
   
		\begin{algorithmic}[1] 
			\STATE  {\bf{for}} $ j \in \mathcal{J} $ 
				\STATE  {\bf{for}} $ i \in {\bf{\bar{N}}}_j  $ 
			\STATE   Sort vehicles based on difference average speed in increasing order, and assign weights $\{{\bf{w}}_s\}$ to them
			\STATE Sort vehicles based on the number of neighbors in descending order,  and assign weights $ \{{\bf{w}}_{n}\}$ to them
			\STATE  Sort vehicles based on residual path in ascending order, assign weight $\{{\bf{w}}_p\}$ to them
		
			\STATE  Calculate a weighted cumulative value for each vehicle $v_i$, as:
			$w_i = w_{d_i} . v_{d_i} + w_{n_i} . \eta_{i} + w_{p_i} . \xi_i$ 
			\STATE Set $\{ {\bf{CB}}_j\}=\{ {\bf{CB}}_j\} + v_i $
						\STATE Sort vehicles in $\{ {\bf{CB}}_j\} $ based on their $w_i$ in descending order
							\STATE {\bf{end} \bf{for}}
			\STATE {\bf{If}} the current cluster head departs from the cluster
			\STATE Select the alternative cluster head, $CH_j$, from the top of $\{ {\bf{CB}}_j\} $
			\STATE Set $\{ {\bf{CB}}_j\}=\{ {\bf{CB}}_j\} - CH_j $
		\STATE	{\bf{end} \bf{if}}
			\STATE {\bf{end} \bf{for}}

		\end{algorithmic}
	\end{algorithm}

	\indent Once the cluster head is selected and successfully communicates with the UAV, the designated CH sends its beacon message containing the ID of the connected vehicles to the UAV at regular intervals. If the UAV fails to receive the beacon message from the CH, it interprets it as the CH leaving the cluster. In such a scenario, the UAV collecting all the CAMs investigates whether there are any unconnected vehicles present within the cluster. If such vehicles are detected, the UAV then proceeds to select an alternate CH from the prioritized backup list.
	
	\subsection{Cluster Head Backup List Scheme}
Immediately following selecting the CH, the UAV employing the Analytic Hierarchy Process (AHP), provides a proficient backup list of capable vehicles suitable for an alternative CH \cite{darko2019review}. To provide this list, the UAV explores the remaining vehicles from Algorithm 2 and subsequently sorts these vehicles in three sequences based on average speed difference, number of neighbors, and residual path within the cluster. The UAV then assigns certain weights to these sorted sequences. Cumulative values of all relevant parameters are calculated for each vehicle and a numerical value is assigned to each.\\
\indent This compiled list is designed to serve as a resource for selecting an alternative CH whenever the current one leaves the cluster. The proposed CH backup selection approach is depicted in Algorithm 3.


	\section{Numerical Results}
	In this section, we investigate the cluster stability performance of the proposed
	scheme in terms of the total number of leaving and re-selection CHs and the obtained SNR between CHs and co-cluster vehicles. For a fair evaluation, we consider other specific benchmark schemes such as random CH selection, and VMaSC \cite{bench} which selects the CH based on the lowest average relative speed. We consider a suburban environment with $3$ UAVs and 12 vehicles traveling at speeds between $40$ - $60$ km/h along a two-lane two-way road of $1$ km length. We also assume that the clustering algorithm is performed every $70$ seconds, while the vehicles send their CAM every 10 seconds. Moreover, the transmission power of vehicles and noise power are considered to be $-70$ and $-114$ dBm, respectively. We also consider the path loss at a reference distance of $1$ m to be $10^{-5}$. 
	Simulations are carried out via MATLAB using a computer with Intel(R) Core(TM) i7-12700 and 32 GB RAM.

 		\begin{figure} [t]
		\centering
		\includegraphics[width=78mm, height=65mm, keepaspectratio, keepaspectratio]{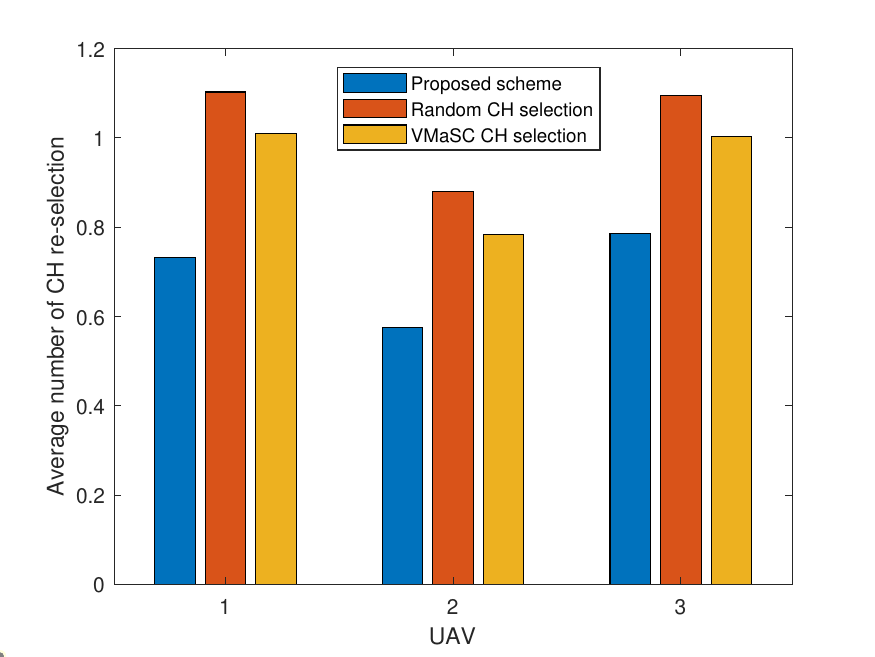}
		\caption{Average number of CHs re-selections per UAV for different schemes } \label{fig3} 
	\end{figure}

		\begin{figure} [t]
		\centering
		\includegraphics[width=78mm, height=65mm,keepaspectratio]{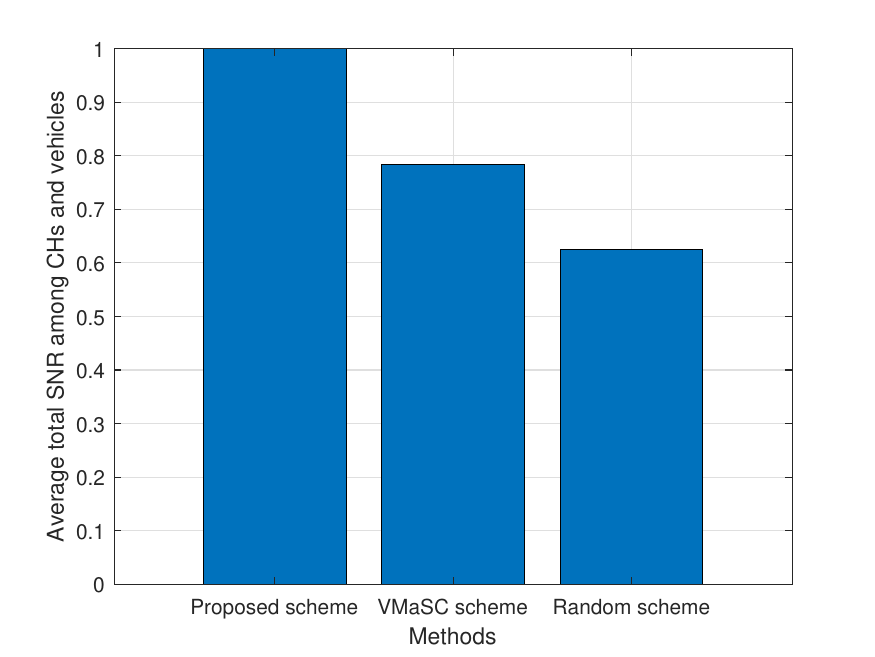}
		\caption{Normalized average total SNR among the selected CHs and co-cluster vehicles} \label{fig4} 
	\end{figure}
	\begin{figure} [t]
		\centering
		\includegraphics[width=78mm, height=65mm, keepaspectratio]{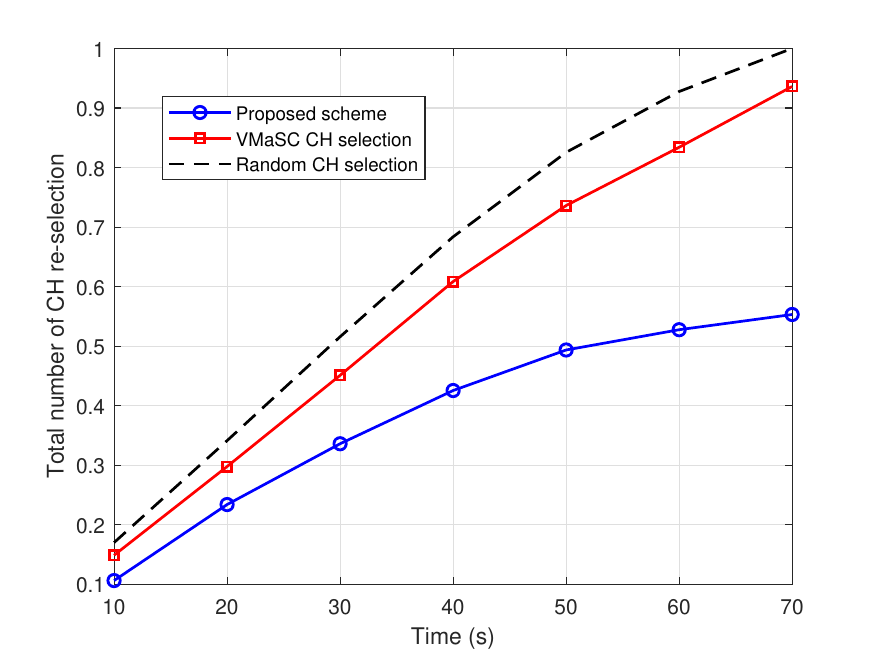}
		\caption{Normalized total number of CHs re-selections versus time } \label{fig5} 
	\end{figure}
			\begin{figure} [t]
			\centering
			\includegraphics[width=78mm, height=65mm]{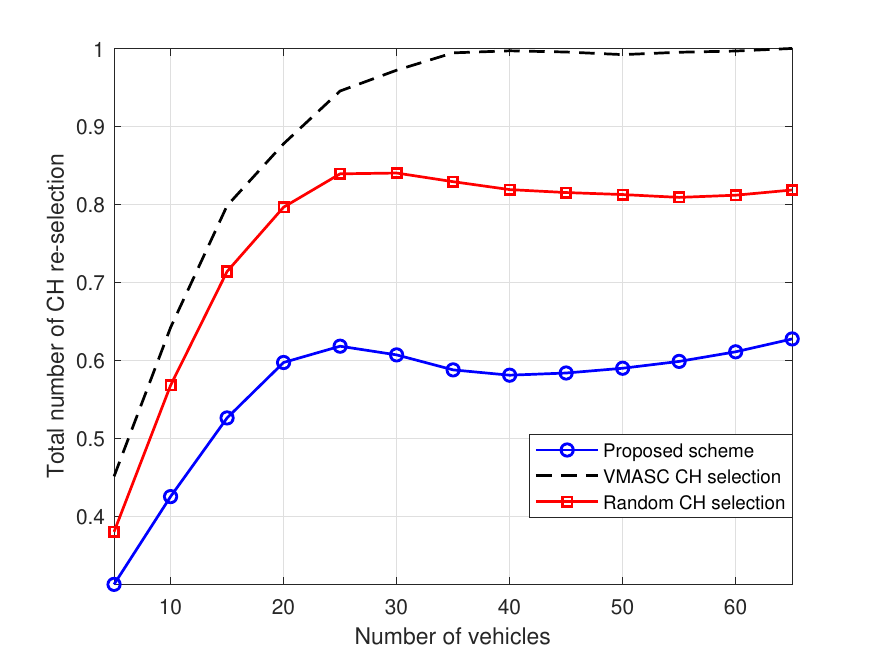}
			\caption{Normalized total number of CHs re-selections versus the number of vehicles} \label{fig6} 
		\end{figure}
  
	Fig. 3 illustrates the average number of CH re-selections for each cluster, comparing the proposed scheme with other benchmarks. The figure shows that for all clusters, the number of CH re-selections in the proposed scheme is less than in the benchmark methods. This indicates that the proposed clustering scheme is more stable than the other methods. The reduced frequency of CH changes means that clusters remain reliable and stable for longer periods. Another observation from the figure is that the number of CH re-selections for the middle cluster is less than for the two side clusters across all schemes. This may be because the middle cluster usually is assigned more vehicles, which allows for more efficient CH selection.
			
		In Fig. 4, we compare the normalized total SNR among CHs and vehicles within clusters for the proposed scheme and two other benchmarks. The total SNR is averaged over the number of CH re-selections and the number of simulations. To ensure a fair comparison, all results are normalized to the maximum obtained SNR. As expected, the proposed scheme outperforms other benchmarks in terms of SNR. The reason is that the proposed scheme considers several criteria such as the average speed, position, and direction of vehicles when selecting a CH. Consequently, the selected CH has a better channel gain with other vehicles compared to other schemes. As a result, the obtained SNR is higher than other schemes with the same transmission power and noise power.
			
			Fig. 5 compares the total number of CH re-selection of the proposed scheme with the other two benchmarks for all clusters as time increases. This comparison metric is the total number of changes for all clusters, normalized by the highest value. It can be seen that as time increases, the total number of CHs leaving the cluster increases due to the movement of vehicles. In clustering schemes, the lower the number of CH re-selections, the higher the cluster stability and the less time spent on selecting an alternative CH. Therefore, it can be observed that the performance of the proposed scheme is better than the benchmarks. One reason is that, considering the collected CAMs, the proposed scheme selects the CH based on vehicle behavior especially based on its average speed and position, which leads to choosing a CH that stays in the cluster longer than other vehicles and leads to cluster stability.
	\begin{table}
\centering
\caption{Clustering Robustness Likelihood}
\label{tab:LinkBudgetPara}
\begin{tabular}{|l|l|}
\hline
\textbf{Scheme}               & \textbf{Likelihood Value} \\ \hline
The proposed scheme             & 0.83            \\ \hline
The VMaSC scheme                   & 0.47             \\ \hline
The random scheme                   & 0.32            \\ \hline
\end{tabular}
\end{table}

	\indent In Fig. 6, the normalized total number of CHs changes of the proposed and two other benchmarks with the change in the number of vehicles is shown. One can see from the proposed scheme that increasing $I$ leads to an increase in the number of CHs changes until $I$=25, after which it remains almost unchanged. This is not far-fetched because, with a small number of vehicles, the number of clusters is also small, reducing the need to select a cluster head. As the number of vehicles increases, each cluster must elect a CH and subsequently the alternative CH. However, with a further increase in the number of vehicles, the number of cluster head changes does not change much. Also, the advantage of the proposed scheme in terms of the number of CH re-selections is demonstrated.\\	
	\indent Finally, Table I represents a clustering robustness comparison of the proposed scheme against the other two benchmarks. As a fair and comprehensive evaluation based on the obtained numerical results, we introduce the clustering robustness likelihood. This metric combines the normalized results of cluster head change rate and SNR with weights $w_R=0.6$ and $w_S=0.4$, respectively as $L=\exp\big(-({{w_R\lambda_R+w_S \lambda_S}})\big)$, where $\lambda_R$ and $\lambda_S$ are the contributions of CH re-selections and SNR to the likelihood, respectively. Given the normalized value of CH re-selections, $R$, and SNR, $S$, while considering Poisson distribution for CH re-selections and Gaussian distribution for SNR, the clustering robustness likelihood can be obtained for our proposed scheme and benchmarks\footnote{We consider Poisson distribution for CH re-selections with parameter $\lambda_r=0.5$, and Gaussian distribution for SNR with parameters $\mu=1$, $\sigma^2=0.1$. Therefore, $\lambda_R=-\ln \big(P(R|\lambda_r)=\frac{\lambda_r^Re^{-\lambda_r}}{R!}\big)$, and $\lambda_S=-\ln \big(P(S|\mu,\sigma^2)=\frac{1}{\sqrt{2\pi\sigma^2}}\exp{(-\frac{(S-\mu)^2}{2\sigma^2})}\big)$ \cite{Like}.}. Based on this analysis, Table I demonstrates that the proposed scheme not only achieves a high robustness likelihood but also significantly outperforms the two benchmark schemes in terms of robustness. Meanwhile, our future work, developing the proposed scheme may study and evaluate clustering robustness likelihood in game theory and machine learning-based clustering algorithms \cite{R2}.
		
		\vspace{-0.2mm}\section{CONCLUSION}
		In this paper, we proposed a new scheme including clustering, selecting a stable cluster head, and providing a cluster head backup list for vehicular communication supported by UAVs. Using a weighted approach, the proposed CH selection algorithm considers four important criteria of vehicle behavior to address all the parameters required to increase the stability of the dynamic network. These criteria are the average speed, position, direction, and number of neighboring vehicles, which lead to the selection of a CH that stays much longer than others within the cluster and can also establish good communication links with vehicles in the same cluster. In addition, providing a backup list leads to an effective reduction of the new CH selection time. Through numerical evaluation, we demonstrated that our proposed clustering scheme improves the reliability and stability of the clusters. This is evidenced by the significantly lower number of cluster heads leaving their clusters compared to other benchmark schemes. We also showed that the proposed scheme is not much affected by the change in the number of vehicles, which ensures its efficiency in dense networks where resource constraints pose significant challenges. As part of our future work, we propose UAVs trajectory design and resource optimization for different clusters as a hybrid cooperative and non-cooperative game between clusters. We also plan to develop the proposed scheme considering real-time vehicle behavior monitoring by minimizing the time required to complete the clustering and CH selection processes.

		
		\ifCLASSOPTIONcaptionsoff
		
		\fi

		
	\end{document}